\documentclass[12pt]{article}
\usepackage[dvips]{epsfig}
\usepackage{amsmath}

\begin{document}

%
\newcommand{\be}{ \begin{equation}}
\newcommand{\ee}{ \end{equation}  }
\newcommand{\bea}{ \begin{eqnarray}}
\newcommand{\eea}{ \end{eqnarray}  }
\newcommand{\bi}{\bibitem}
\newcommand{\rd}{ \mbox{\rm d} }
\newcommand{\rD}{ \mbox{\rm D} }
\newcommand{\re}{ \mbox{\rm e} }
\newcommand{\rO}{ \mbox{\rm O} }
\newcommand{\erf}{\mbox{\rm erf}}
\newcommand{\diag}{\mbox{\rm diag}}

\renewcommand{\floatpagefraction}{0.8}

\def\del{\partial}

\def\bbar#1{\bar{\bar{#1}}}
\newcommand{\tauint}[1]{\tau_{{\rm int}, #1 }}
\newcommand{\bbtauint}[1]{\bbar{\tau}_{{\rm int}, #1 }}

\title{
Monte Carlo errors with less errors.
}

\author{
Ulli Wolff\thanks{
e-mail: uwolff@physik.hu-berlin.de} \\
Institut f\"ur Physik, Humboldt Universit\"at\\ 
Newtonstr. 15 \\ 
12489 Berlin, Germany
}
\date{}
\maketitle
\vbox{
\centerline{
\epsfxsize=2.5 true cm
\epsfbox{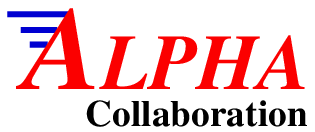}}
}
\vskip 0.5 cm

\vspace*{3cm}

\begin{abstract}
We explain in detail
how to estimate mean values and assess statistical errors
for arbitrary functions of elementary observables in Monte Carlo
simulations.
The method is to estimate and sum the relevant autocorrelation
functions, which is argued to produce more certain error estimates
than binning techniques 
and hence to help toward a better exploitation of expensive simulations.
An effective integrated autocorrelation time is computed which is suitable
to benchmark efficiencies of  simulation algorithms with regard to
specific observables of interest. A Matlab code is offered for download 
that implements the method. It can also combine independent runs (replica) allowing
to judge their consistency.

\end{abstract}
\begin{flushright} HU-EP-03/32 \end{flushright}
\begin{flushright} SFB/CCP-03-12 \end{flushright}
\thispagestyle{empty}
\newpage


\section{Formulation of the problem}

In this article we collect some theory
and formulas for the practical estimation of statistical and
systematic errors in Monte Carlo simulations.
Emphasis is put on the estimation of in general nonlinear functions
(`derived quantities') of the primary expectation values.
The strategy focuses on the explicit determination of 
the relevant autocorrelation
functions and times. 
We shall discuss why this is advantageous
compared to the popular binning techniques, which handle autocorrelations
only implicitly.
A Matlab code is made available, which implements the method described here.

The material given here is only partially new.
It has been briefly discussed in the appendix of \cite{Frezzotti:2000rp}
and in internal notes of the ALPHA collaboration \cite{ALPHAMCerr}
and in lecture notes \cite{CP2}.
However, due to the fact that mastering this topic
is an indispensable  prerequisite
for every student or researcher in the popular field of
lattice simulation, we felt that there is a
gap in the readily citeable literature which we want to fill here.


We  assume to have a number of primary observables enumerated
by a Greek index with exact statistical mean values $A_\alpha$.
The object to be estimated is a function of these,
\be
F \equiv f(A_1,A_2,\ldots) \equiv f(A_\alpha).
\label{Fdef}
\ee
A typical example is given by reweighting, where a quotient
$F=A_1/A_2$ has to be determined. A more involved case are fit-parameters
determined by correlation functions over a range of separations.

For the estimation we employ Monte Carlo estimates
of the primary observables given by $a_\alpha^i$. 
For each observable there are $i=1,2,\ldots,N$ successive estimates
separated by executions of some valid update procedure.
We assume that the Markov chain has been
equilibrated before recording data beginning with $i=1$. 
This has to be secured from case to case,
and we in particular recommend visual inspections of initial
histories of observables to decide on safely discarding data.
A posteriori one should also check that the equilibration
time was much longer than autocorrelation times found in the
analysis of the supposed equilibrium data.

A key theoretical quantity for error estimation is the 
autocorrelation \cite{Madras:1988ei,SokalCours}  function
$\Gamma_{\alpha\beta}$ defined by
\be
\left\langle (a_\alpha^i - A_\alpha) (a_\beta^{i+n} - A_\beta) \right\rangle
= \Gamma_{\alpha\beta}(n).
\label{Gammadef}
\ee
that correlates the deviation of the $i$'th estimate for $A_\alpha$
with the deviation for variable $\beta$ after performing $n \ge 0$ updates.
Averages $\langle \,\cdot\,  \rangle$ here and {\bf throughout this paper}
refer to an ensemble of identical numerical
experiments with independent random numbers and initial states
--- a concept useful
for errors that are themselves statistical in nature.
The dependence of $\Gamma_{\alpha\beta}$ on the separation $n$ alone
is due to being in equilibrium.

We take the simulation to produce configurations $\phi_i$ distributed with the
normalized Boltzmannian $P(\phi)$.
The update algorithm is specified by transition probabilities
$W(\phi \to \phi')$ and the measurements by $a_\alpha^i = {\cal O}_\alpha(\phi_i)$.
Then the autocorrelation function in equilibrium is given as a double sum over
configurations\footnote{
We here use a discrete notation for sums over configurations. The generalization
to continuous fields with integrations is obvious.
},
\be
\Gamma_{\alpha\beta}(n) = \sum_{\phi}  \sum_{\phi'} P(\phi) W^n(\phi \to \phi')
({\cal O}_\alpha(\phi)-A_\alpha) ({\cal O}_\beta(\phi')-A_\beta).
\ee
Here $W^n$ characterizes transitions by a succession of $n$ update steps.
If $W$ obeyed detailed balance with respect to $P$, we would conclude
symmetry in the indices $\alpha\beta$. In practice we usually have
only stability and no such symmetry
seems to be implied in general. 
It is however natural and useful to define
\be
\Gamma_{\alpha\beta}(-n) = \Gamma_{\beta\alpha}(n)
\ee
implying a combined symmetry.
Obviously $\Gamma_{\alpha\beta}(0)$ is the symmetric
covariance matrix with the
ordinary variances on the diagonal. It is a (static) property of the simulated
statistical system alone, while $\Gamma$-values for nonzero separation 
are dynamic and depend
on the Monte Carlo algorithm in use.

We end this theoretical setup by a slight generalization
of our analysis. We add another index to the  estimates
of the primary observables, $a_\alpha^{i,r}$. 
The index $r=1,2,\ldots,R$
counts a number of statistically independent replica. These
are independent simulations which typically arise from repeated
runs, or in particular,
if parallel computers are used in the `trivial parallelization' 
mode\footnote{The reader is invited to simplify all formulas to $R=1$
in a first reading.}.
For each replicum $r$ there are $i=1,2,\ldots,N_r$ successive 
equilibrium estimates
separated by executions of {\em the same update procedure}.
Obviously (\ref{Gammadef}) becomes
\be
\left\langle (a_\alpha^{i,r} - A_\alpha) (a_\beta^{j,s} - A_\beta) \right\rangle
= \delta_{rs} \Gamma_{\alpha\beta}(j-i).
\label{Gammadefr}
\ee

\section{Errors and biases}

\subsection{Primary observables\label{primobs}}

We define the per replicum means
\be
\bar{a}_\alpha^r = \frac1{N_r} \sum_{i=1}^{N_r} \; a_\alpha^{i,r}.
\ee
and in terms of them
the natural estimator for the primary quantities $A_\alpha$
\be
\bbar{a}_{\alpha} = \frac1{N} \sum_{r=1}^R \; N_r \, \bar{a}_\alpha^r
\ee
with the total number of estimates
\be
N = \sum_{r=1}^R N_r \, .
\ee
Introducing the deviations
\be
\bar{\delta}_\alpha^r = \bar{a}_\alpha^r - A_\alpha\, , \quad
\bbar{\delta}_\alpha = \bbar{a}_\alpha - A_\alpha \, .
\ee
we state that these estimators are unbiased,
\be
\langle \bar{\delta}_\alpha^r  \rangle = 0 =
\langle \bbar{\delta}_\alpha  \rangle.
\ee
We assume a normal distribution\footnote{
Note that the $a_\alpha^{i,r}$ themselves 
need not be Gaussian distributed and that large
$N_r$ makes the $\bar{a}_\alpha^{r}$ Gaussian by the central limit theorem.}
which is then completely defined by the covariance matrix given by
\be
\langle \; \bar{\delta}_\alpha^r \, \bar{\delta}_\beta^s \; \rangle
= \frac1{N_r^2} \sum_{i,j=1}^{N_r} \Gamma_{\alpha\beta}(j-i)\delta_{rs}
= \frac1{N_r} C_{\alpha\beta} \, \delta_{rs} \times(1+ \rO(\tau/N_r)) \, 
\label{Cdef}
\ee
where
\be
C_{\alpha\beta} = \sum_{t=-\infty}^{\infty} \Gamma_{\alpha\beta}(t)
\label{CfromGamma}
\ee
does not depend on the run-length.
Here and in the following it is assumed that a finite scale $\tau$
characterizes the asymptotic exponential decay of
$\Gamma_{\alpha\beta}$ with $|t|$.
The claimed $N_r$-dependence holds for runs with lengths $N_r \gg \tau$
and if this condition is violated, an error estimation is hardly possible.
For the covariance matrix of $\bbar{a}_\alpha$
\be
\langle \; \bbar{\delta}_\alpha \, \bbar{\delta}_\beta \; \rangle
= \frac1{N} C_{\alpha\beta} \times(1+ \rO(R\tau/N))
\ee
follows.
Roughly speaking, as expected,
$\bbar{a}_\alpha$ differs from $A_\alpha$ by 
an error of order $1/\sqrt{N}$ and 
the task of an error analysis amounts to a reasonably
accurate estimation of $C_{\alpha\beta}$ which includes
all autocorrelation effects.

\subsection{Derived quantities}
To determine $F$ we consider estimators
\be
\bbar{F} = f(\bbar{a}_\alpha)
\label{Fbb}
\ee
and
\be
\bar{F} = \frac1{N} \sum_{r=1}^R \, N_r \, f(\bar{a}_\alpha^r).
\label{Fbardef}
\ee
We assume the estimates of the primary observables to be accurate enough
to justify Taylor expansions of $f$ in the fluctuations, for example
\be
\bbar{F} = F + \sum_\alpha f_\alpha \bbar{\delta}_\alpha 
+\frac12 \sum_{\alpha\beta}  f_{\alpha\beta}\, \bbar{\delta}_\alpha  \bbar{\delta}_\beta
+\dots
\ee 
with derivatives
\be
f_\alpha = \frac{\del f}{\del A_\alpha}, \quad 
f_{\alpha\beta} = \frac{\del^2 f}{\del A_\alpha \del A_\beta}
\label{fder}
\ee
{\em taken at the exact values} $A_1, A_2, \ldots$.
It follows that our estimators for $F$ are biased unless $f$ is
linear,
\be
\langle \bbar{F} - F \rangle  \simeq
\frac1{2N} \sum_{\alpha\beta}  f_{\alpha\beta}\, C_{\alpha\beta} 
\simeq \frac1R  \, \langle \bar{F} - F \rangle \, ,
\ee
Usually, this bias is negligible compared to statistical errors
due to large enough $N$. Replica ($R\ge2$) can however be used to control
and, if deemed to be appropriate, cancel the leading bias by using
\be
\langle \bbar{F} - F \rangle \simeq \frac1{R-1} \, \langle \bar{F} - \bbar{F} \rangle
\ee
to replace
\be
\bbar{F} \to \frac{R \bbar{F} - \bar{F}}{R-1} \, .
\label{unbias}
\ee

The error $\sigma_F$ is to leading order given by
\be
\sigma_F^2 = \langle\, (\bbar{F} - F)^2 \,\rangle \simeq  
\frac1{N} C_F 
\label{sigmaF}
\ee
with
\be
C_F = \sum_{\alpha\beta}  f_\alpha f_\beta \, C_{\alpha\beta} \, .
\ee
We rewrite this expression as
\be
\sigma_F^2 = \frac{2 \tauint{F} }{N} \, v_F
\label{sigbytauint}
\ee
with the effective `naive' -- i.e. disregarding autocorrelations --
variance relevant for $F$
\be
v_F = \sum_{\alpha\beta} f_\alpha f_\beta \Gamma_{\alpha\beta}(0)
\ee
and the integrated autocorrelation time for $F$
\be
\tauint{F} = \frac1{2 v_F} 
\sum_{t=-\infty}^{\infty} \sum_{\alpha\beta} f_\alpha f_\beta \Gamma_{\alpha\beta}(t).
\label{tauintF}
\ee
This one number
encodes the efficiency of the algorithm in use
for a determination of $F$, if its execution time per update is known.
To appreciate this notation for the ratio
of the true to the naive error
two extremal cases are instructive. In the absence of
autocorrelations, $\Gamma_{\alpha\beta}(t) \propto \delta_{t,0}$,
we have $2 \tauint{F}=1$ and (\ref{sigbytauint}) is the standard error.
For a purely exponential behaviour, 
$\Gamma_{\alpha\beta}(t) \propto \exp(-|t|/\tau)$,
this scale is recovered, $\tauint{F} = \tau +\rO(\tau^{-1})$.
In general there are however many contributions and $\tauint{F}$ is
more like a `dynamical susceptibility'. Another interpretation of
(\ref{sigbytauint}) is, that only the reduced number of $N/(2\tauint{F})$ 
estimates are effectively
independent to bring down statistical errors.

If we have $R \ge 2$ replica another consistency check is useful.
The replica estimates $f(\bar{a}_\alpha^r)$ are assumed to be normally
distributed. Their variances are
\be
\left\langle\, \left(f(\bar{a}_\alpha^r)-F\right)^2 \,\right\rangle
= \frac{C_F}{N_r} .
\ee
We may imagine to do a $\chi^2$ fit of the replica estimates to a constant $K$
by minimizing
\be
\chi^2(K) = \sum_r \frac{(f(\bar{a}_\alpha^r)-K)^2}{C_F/N_r}.
\ee
The minimum occurs precisely at $K=\bar{F}$  given in (\ref{Fbardef}).
The probability to find in such a `fit' a minimal $\chi^2 \ge x$ is
given by
\be
Q = 1 - P(x/2,(R-1)/2),
\label{Qdef}
\ee 
where $P$ is the incomplete Gamma function,
\be
P(U,n) = \frac1{\Gamma(n)} \int_0^{U} du\, u^{n-1}\, \exp(-u).
\ee
Taking the actually observed $x=\chi^2(\bar{F})$ we obtain
the so-called goodness of fit.
It is very unlikely to see
$Q$-values much less than 0.1, if the underlying
distribution has the required properties, and this is hence a good
consistency check \cite{Numrec}.
In a histogram plot we may in addition inspect
\be
p_r = \frac{f(\bar{a}_\alpha^r)-\bar{F}}{\sigma_F \sqrt{N/N_r-1}}
\label{repdist}
\ee
which should have zero mean and unit variance. 

In the next section 
we develop numerical estimators for $v_F$ and $\tauint{F}$,
which at present are not yet of practical use, since they
so far involve the numerically unknown exact $A_\alpha$ and $\Gamma_{\alpha\beta}$.

\section{Extraction of errors from measured data}

\subsection{Error estimators}

We now describe what we would like to call the $\Gamma$-method
of error estimation (as opposed to binning, see App.\ref{appbinning}),
where we explicitly estimate the autocorrelation function.

In terms
of the given estimates $a_\alpha^{i,r}$
we define an estimator $\bbar{\Gamma}_{\alpha\beta}$ 
of the autocorrelation function
\be
\bbar{\Gamma}_{\alpha\beta}(t) = \frac1{N-Rt} \sum_{r=1}^R
\sum_{i=1}^{N_r-t}
(a_\alpha^{i,r} - \bbar{a}_\alpha)  (a_\beta^{i+t, r} -\bbar{a}_\beta)
\label{Gammaest}
\ee
where $0 \le t \ll N_r$ for all $r$ is understood.
An important principle underlying this choice is to avoid
unnecessary noise from terms which vanish (only) on average.
Therefore
the correlation is first taken within each replicum and only then averaged.
Using arguments of the same type as for (\ref{Cdef}) 
the leading bias of this estimator is evaluated,
\be
\langle\, \bbar{\Gamma}_{\alpha\beta}(t) \,\rangle - \Gamma_{\alpha\beta}(t)
\approx - \frac{C_{\alpha\beta}}{N} .
\label{Gammabias}
\ee
It comes from subtracting in (\ref{Gammaest}) the ensemble mean instead of the
exact average. For uncorrelated data this is the well known bias
in the estimator of the variance cancelled by the  $N-1$ instead of $N$
normalization. For the moment we neglect this bias of order $1/N$ relative
to the  $1/\sqrt{N}$ statistical error in $\bbar{\Gamma}_{\alpha\beta}$
to be worked out in the next subsection, 
and come back to it later. 

For the derived quantity $F$ only the projected autocorrelation is required,
\be
\bbar{\Gamma}_F(t) = \sum_{\alpha\beta} \bbar{f}_\alpha \bbar{f}_\beta
\bbar{\Gamma}_{\alpha\beta}(t)
\ee
where $\bbar{f}_\alpha$ is the gradient as in (\ref{fder}) but
now evaluated at arguments $\bbar{a}_1, \bbar{a}_2, \ldots$,
which inflicts a small relative error of order $1/N$ that we do not attempt
to cancel.

As practical estimators for $v_F$ and $C_F$ we take
\be
\bbar{v}_F = \bbar{\Gamma}_F(0)
\ee
and
\be
\bbar{C}_F(W) = 
\left[ \bbar{\Gamma}_F(0) + 2\sum_{t=1}^W \bbar{\Gamma}_F(t) \right] \, .
\label{CFbbarDef}
\ee
A bias results from the truncation of the autocorrelation sum,
\be
\frac{\langle\, \bbar{C}_F(W)\,\rangle - C_F}{C_F}  
\sim  - \exp(-W/\tau).
\label{syserrW}
\ee
For a purely exponential $\Gamma_F$ this is an approximate equality,
otherwise an order of magnitude statement with the scale $\tau$ introduced
at the end of section \ref{primobs}.
The summation window $W$ has to be chosen with care.
On the one hand it has to be large compared to the decay time $\tau$
for a small systematic error, on the other hand with too large $W$
one includes terms with negligible signal but excessive noise.
In the next subsection we shall define an automatic procedure
to choose $W$.

In a frequently met case we have many primary observables $A_\alpha$
but only want to estimate a few functions of them. 
Then it is advantageous to actually
not compute the full autocorrelation matrices
$\bbar{\Gamma}_{\alpha\beta}(t)$ but to just 
form $\bbar{a}_\alpha$ and $\bbar{f}_\alpha$ and
immediately project
the measured data
\be
a_\alpha^{i,r} \to a_f^{i,r} = \sum_\alpha \bbar{f}_\alpha a_\alpha^{i,r}
\ee
to base the remaining autocorrelation analysis directly on them.

Another practical point concerns the computation of $\bbar{f}_\alpha$.
It may well be inconvenient or even impossible to analytically
specify the gradient of $f$ and we may prefer to evaluate it numerically
based on the supplied routine for $f$ itself only.
A natural scale
$h_\alpha$ is extracted
from the data,
\be
h_\alpha = \sqrt{\frac{\bbar{\Gamma}_{\alpha\alpha}(0)}{N}} \; ,
\ee
and we take 
\be
\bbar{f}_\alpha \approx \frac1{2h_\alpha} 
[f(\bbar{a}_1,\bbar{a}_2, \dots ,\bbar{a}_\alpha + h_\alpha, \dots) 
-f(\bbar{a}_1,\bbar{a}_2, \dots ,\bbar{a}_\alpha - h_\alpha, \dots)]
\ee
as a numerical estimate for the gradient of $f$ with negligible errors of 
$\rO(h_\alpha^2) \propto 1/N$.

\subsection{Error of the error}

When an error is assessed numerically in a Monte Carlo
it is itself affected
by statistical error. This is usually not quantified. But to
rate algorithms on the basis of the $\tauint{F}$ they produce
for physically interesting $F$ it is better to at least have a rough idea.
This is even more true, if one wants to falsify a theoretical
prediction based on
one of these `$2.5 \, \sigma$ discrepancies', while maybe the estimate of $\sigma_F$ itself
is uncertain by a factor 2.

In appendix \ref{Sokalerror} the very convenient approximate
formula 
\be
\langle (\bbar{C}_F(W)-C_F)^2 \rangle
\approx  \frac{2(2W+1)}{N} C_F^2 \, .
\label{Sokalformula}
\ee
is derived that was to our knowledge
first given in \cite{Madras:1988ei}.
For the quotient
\be
\bbtauint{F}(W) = \frac{\bbar{C}_F(W)}{2 \bbar{v}_F}
\ee
we similarly find\footnote{
We here use error propagation
and the upper bound 
in (\ref{uFdef}). 
This leads to an upper bound on the error of
$\bbtauint{F}$ regarding the terms beyond the one
$\propto W$ which is both exact and dominant.
} 
with the help of (\ref{errerrvC}) and (\ref{uFdef})
\be
\langle (\bbtauint{F}(W) - \tauint{F})^2 \rangle
\approx \frac{4}N\,(W+1/2-\tauint{F}) \tauint{F}^2 .
\label{Sokalfortau}
\ee

\begin{figure}
\begin{center}
\epsfig{file=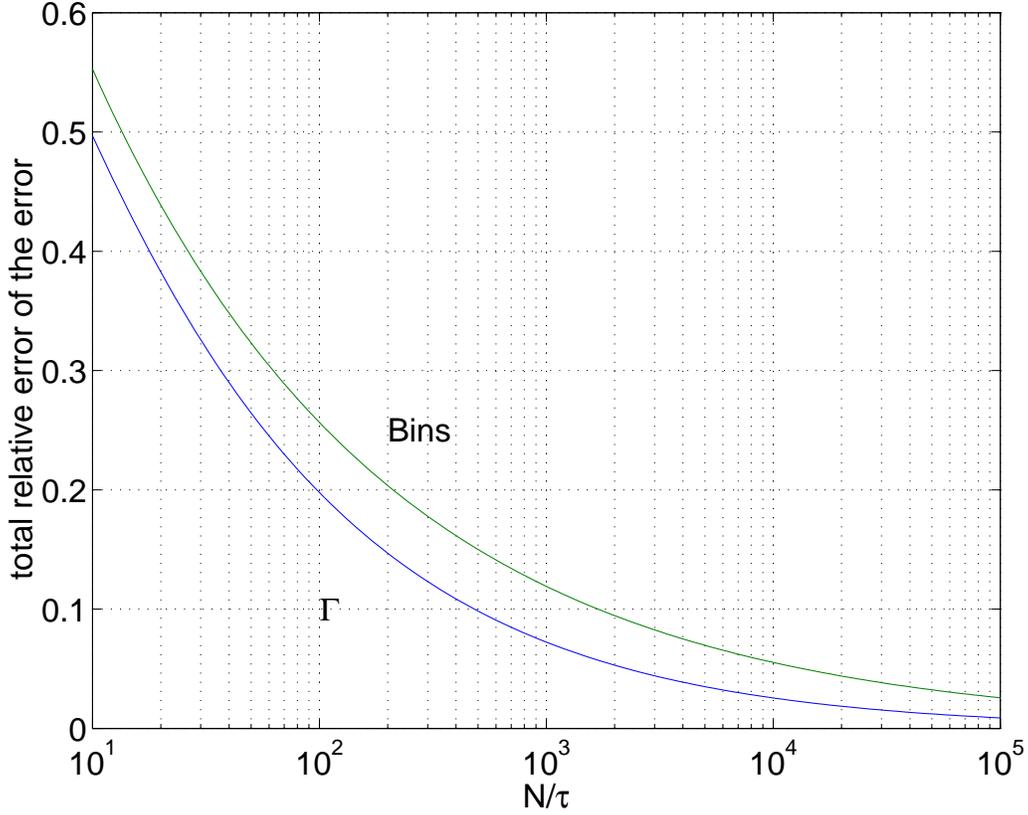,width=\textwidth}
\end{center}
\caption{Relative total error of the error estimate for the $\Gamma$-method
and for binning.}
\label{errerr}
\end{figure}

Beside the statistical error of $\bbar{C}_F(W)$ there is the systematic
error from the $W$-truncation
in (\ref{syserrW}). 
We define as optimal the $W$ value that minimizes the sum
of the absolute values of these errors and yields a total
relative error
\be
\frac{\delta_{\rm tot}(\bbar{\sigma}_F)}{\bbar{\sigma}_F} \approx 
\frac12 \min_W \left(
\exp(-W/\tau) + 2 \sqrt{W/N}
\right)
\label{Gammamin}
\ee
for the final error estimate
\be
\bbar{\sigma}_F^2 = \frac{\bbar{C}(W)}{N}
\ee
evaluated  at the optimal $W$.
It is
a function of $N/\tau$  and plotted  in Fig.\ref{errerr}
for relevant arguments. 
The minimum leads to a simple transcendental equation that we solved numerically, 
but it can
also be well approximated by
$W=\tau \ln(N/\tau)/2$.
The asymptotic behaviour is $\delta_{\rm tot} \propto \sqrt{\ln(N)/N}$. Due
to the exponential behaviour (\ref{syserrW}) the systematic component of the error
becomes negligible at large N, albeit only slowly,
\be
\frac{\delta_{\rm sys}(\bbar{\sigma}_F)}{\delta_{\rm stat}(\bbar{\sigma}_F)}
= \frac{\tau}{2W} \approx \frac1{\ln(N/\tau)} \, .
\ee

A simple and popular alternative to the analysis method just described
are binning methods, where data are pre-averaged over sections of the history
of length $B$. These bin averages are taken as independent estimates.
For both ordinary and jackknife binning  (see appendix \ref{appbinning}
for more details)
a relative 
systematic error of the error estimate occurs, which decays only
proportional to $\tau/B$.
The relative statistical error of the error due to the finite number of bins 
is $\sqrt{2B/N}$.
By balancing these errors similarly as before we find the total error
at optimal bin size
\be
\frac{\delta_{\rm tot}(\bar{\sigma}_{F{\, \rm bin}})}{\bar{\sigma}_{F{\, \rm bin}}} 
\approx \frac12 \min_B \left(
\tau/B + \sqrt{2B/N}
\right) = \frac32 (2N/\tau)^{-1/3}
\ee
which is reached for 
\be
B=\tau (2N/\tau)^{1/3}
\label{Bopt}
\ee
and is also shown in Fig.\ref{errerr}.
Here the ratio of systematic to statistical error is constant,
\be
\frac{\delta_{\rm sys}(\bar{\sigma}_{F{\, \rm bin}})}{\delta_{\rm stat}(\bar{\sigma}_{F{\, \rm bin}})}
= \frac12 \, .
\ee

In comparing the $\Gamma$-method with binning it is clear now that
the bin size $B$ and the window size $2W$ play a very similar role.
All advantages of the $\Gamma$-method derive from the exponential in $W$
rather than inverse power in $B$ behaviour of the systematic error of the error.
One could say, that in this way we have  an improved estimator of 
the error whose own error
decays with a (slightly) larger power of $N$.

We briefly come back to the bias in our error estimate
caused by the right hand side of (\ref{Gammabias}). Cancelling this term when
estimating $C_F$ by (\ref{CFbbarDef}) amounts to enlarging
\be
\bbar{C}_F(W) \to \bbar{C}_F(W) \left(1+\frac{2W+1}{N} \right) \, ,
\ee
which at the optimal $W$ value
represents a correction down  by a factor $1/\sqrt{N}$ (up to logs)
relative to both contributions in (\ref{Gammamin}).
Since the correction is simple and known in closed form we include it
in applications.

\subsection{Automatic windowing procedure \label{window}}

We present now an algorithm to automatically choose $W$
that is similar but not identical to the one 
proposed in \cite{Madras:1988ei}.
To determine $W$ in an actual data analysis,
we start from a hypothesis that $\tau \sim S \tauint{F}$ 
with some factor $S$.
More precisely, we solve
\be
2 \bbtauint{F}(W) = \sum_{t=-\infty}^{\infty} \exp(-S|t|/\bbar{\tau}(W))
\ee
for $\bbar{\tau}$,
to get a formula that is more precise for small $\bbar{\tau}$,
\be
\frac{S}{\bbar{\tau}(W)} =
\ln\left(\frac{2\bbtauint{F}(W)+1}{2\bbtauint{F}(W)-1}\right)
= \frac1{\bbtauint{F}} + \frac1{12 \, \bbtauint{F}^3} + \ldots .
\ee
If $\bbtauint{F} \le 1/2$, we set $\bbar{\tau}(W)$ to a tiny
positive value.
Then we
compute
for $W=1,2,\ldots$
\be
g(W) = \exp[-W/\bbar{\tau}(W)] - \bbar{\tau}(W)/\sqrt{WN} \, .
\ee
Up to a factor this is
the $W$-derivative defining the minimum in (\ref{Gammamin}).
The first value $W$ where $g(W)$ is negative (sign change)
is taken as our summation window
which under our hypothesis is self-consistently close to optimal.
For most practical systems, 
algorithms and observables $\tau$ and $\tauint{F}$
are of the same order, and $S=1\ldots 2$ is a reasonable choice. A too large $S$
leads to (in principle)
unnecessarily large statistical errors of the error.

It is important to verify by eye, at least for a representative set of
observables, that $\bbtauint{F}(W)$ with
its statistical errors (\ref{Sokalfortau}) 
as a function of $W$ exhibits a plateau around the 
automatically chosen value.
If this is not satisfactory, one has
to modify $S$ to achieve this.
A similar procedure could be set up
to semi-objectively choose the bin size $B$ although
this will be less stable, since the plateau behaviour
will be less pronounced.

\section{Conclusions}

We have given a detailed description of the error analysis of
Monte Carlo data. We found advantages in explicitly analyzing
autocorrelation functions rather than employing binning strategies.
They are due to a component in the systematic error of the error estimate
from binning which is proportional to $\tau/B$. It is a finite
bin-size effect in  the integrated autocorrelation time
from these procedures, which is avoided in the $\Gamma$ method.
The discussion includes a careful analysis of the errors of the
error estimate, both statistical and systematic, and a minimization
of their sum. Their approximate size can be read off from Fig.\ref{errerr}
once one has an at least rough estimate of the autocorrelation decay scale $\tau$.

A Matlab routine that implements the whole method, including 
the necessary plots,
is described
and offered for download in the internet. It is also capable
of handling multiple runs (or subdividing one run into several parts)
and to judge the compatibility of the multiple estimates relative to
the over-all error estimate.
The routine adapts itself mostly automatically to
the autocorrelation times involved. This works under
reasonable assumptions mentioned in the text, 
but an additional visual inspection
of the autocorrelations for at least part of the observables is
always recommended.

{\bf Acknowledgements.} I would like to thank the members of ALPHA for
discussions and numerous tests of the software described here and its
precursors. Philippe de Forcrand, Andreas J\"uttner, Bj\"orn Leder, Francesco Knechtli, 
Thomasz Korzec, Martin L\"uscher, Juri Rolf and Rainer Sommer helped with feedback 
on the manuscript. 

\begin{appendix}
\section{The Madras-Sokal formula for the statistical error of the error 
\label{Sokalerror}}

To prove (\ref{Sokalformula}) we need to investigate
\bea
&& \langle \, \bbar{\Gamma}_{\alpha\beta}(s) 
\bbar{\Gamma}_{\gamma\delta}(t) \, \rangle 
=  \\ \nonumber
&&
\frac1{N-s}\sum_{i=1}^{N-s} 
\frac1{N-t}\sum_{j=1}^{N-t}
\langle \,
(a_\alpha^{i} - \bar{a}_\alpha)
(a_\beta^{i+s} - \bar{a}_\beta)
(a_\gamma^{j} - \bar{a}_\gamma)
(a_\delta^{j+t} - \bar{a}_\delta)
\, \rangle.
\eea
This formula is for only one replicum at the moment
to avoid even more indices. 
The essential approximation consists now of approximating
this four-point function by its disconnected part,
\bea
&&\langle \, \bbar{\Gamma}_{\alpha\beta}(s) 
\bbar{\Gamma}_{\gamma\delta}(t) \, \rangle 
\approx  (\Gamma_{\alpha\beta}(s)-C_{\alpha\beta}/N) 
         (\Gamma_{\gamma\delta}(t)-C_{\gamma\delta}/N) +
\\ \nonumber
&&
+ \frac1{N^2} \sum_{i,j=1}^N \Bigl[
\Gamma_{\alpha\gamma}(j-i) \Gamma_{\beta\delta}(j-i+t-s) +
\Gamma_{\alpha\delta}(j-i+t) \Gamma_{\beta\gamma}(j-i-s)
\Bigr]
\eea
This is plausible for large $N$. At least gathering
from experience with ordinary statistical systems
the connected part makes a smaller contribution since all
arguments have to be close to each other.
We have also assumed $|s|,|t| \ll N$ to extend the sums
with relative errors of order $|s|/N, |t|/N$.
Furthermore we use for a rapidly decaying function $g$
the approximation
\[
\frac1{N^2} \sum_{i,j=1}^N g(j-i) \approx 
\frac1N \sum_{k=-\infty}^{\infty} g(k)
\]
similarly as in (\ref{CfromGamma}) to obtain
\be
\langle \, (\bbar{\Gamma}_F(s)-\Gamma_F(s)) 
(\bbar{\Gamma}_F(t)-\Gamma_F(t)) \, \rangle
= \frac1N \sum_{k=-\infty}^{\infty}
\Gamma_F(k) [\Gamma_F(k+t-s)+\Gamma_F(k-t-s)].
\ee
Summing now $t,s=-W,\ldots,W$ and assuming
that each factor $\Gamma_F$ cuts off the sum unless its argument is
much smaller in absolute value than $W$ we derive
\be
\langle (\bbar{C}_F(W)-C_F)^2 \rangle \approx
\frac2N (2W+1) \sum_{k,l=-\infty}^{\infty} \Gamma_F(k) \Gamma_F(l)
=\frac2N (2W+1) C_F^2
\label{errerrCC}
\ee
with corrections of $\rO(\tau/W)$.

By very similar steps one may also derive
\be
\langle\, (\bbar{v}_F - v_F) (\bbar{C}_F(W)-C_F) \,\rangle \approx
\frac2N C_F^2.
\label{errerrvC}
\ee
A more complicated case is
\be
\langle\, (\bbar{v}_F - v_F)^2 \,\rangle \approx
\frac2N \sum_{k=-\infty}^{\infty}(\Gamma_F(k))^2
\le \frac2N v_F C_F,
\label{uFdef}
\ee
where the inequality certainly holds if  
$0 \le \Gamma(k) \le \Gamma(0)$ is true. 
Even if this happens to be violated out in the tail of $\Gamma$,
we still expect the inequality to hold for the sum (dominated by small $|k|$)
in practical applications.

Repeating the above analysis with several replica leads to the same 
formulas and thus yields (\ref{Sokalformula}). 
In view of the approximations we note that
the practical experience is,
that the resulting errors of $\tauint{F}$ are consistent with their scatter
when several runs are made.

\section{Binning methods \label{appbinning}}

Here we give the basic formulas for error estimation by binning.
We assume data $a_\alpha^i$ where possible
replica are sewed together to one history of length $N=B N_B$ which we
divide into $N_B$ sections of $B$ consecutive measurements each\footnote{
The breaks of autocorrelation at replica boundaries are neglected and
a possible remainder is discarded.}. 
We form bins or blocked measurements
\be
b_\alpha^k = \frac1{B} \sum_{i=1}^B a_\alpha^{(k-1)B+i} \, ,
k=1,\ldots,N_B
\ee
The mean values are all identical 
$\bar{b}=\bar{a}=\bbar{a}$.
The idea is now to choose $B$ large enough such that the  bins 
$b_\alpha^k$
can be regarded as approximately uncorrelated, 
but small enough such that there remains
a large number $N_B$ of such `events'.
As an  error estimator one then takes 
\be
\bar{\sigma}^2_{F{\, \rm bin}} = \frac{1}{N_B(N_B-1)}
\sum_{k=1}^{N_B} (f(b_\alpha^k) -\bar{F})^2 \, .
\label{sigmabin}
\ee
If we Taylor-expand $f$ again, the relevant correlation matrix is
\bea
&&\left\langle (b_\alpha^k-\bar{a}_\alpha)(b_\beta^k-\bar{a}_\beta)\right\rangle =
\frac1{B^2} \sum_{u,v} \Gamma_{\alpha\beta}(u-v) \\ \nonumber
&& -\frac1{BN} \sum_{u,i} (\Gamma_{\alpha\beta}(u-i)+\Gamma_{\alpha\beta}(i-u))
+\frac1{N^2} \sum_{i,j} \Gamma_{\alpha\beta}(i-j)
\eea
where indices $u,v$ run through bin $k$ only while $i,j$ range
from 1 to $N$ as before. The result is dominated by the first term
\be
\frac1{B^2} \sum_{u,v} \Gamma_{\alpha\beta}(u-v)
\simeq \frac1{B} C_{\alpha\beta} (1 - \tau/B)
\label{syserrB}
\ee
while the other terms approximately sum to $-C_{\alpha\beta}/N$. 
The coefficient of $\tau/B$ is exact if $\Gamma_{\alpha\beta}$ is
simply exponential, but in practice this should be read as O($\tau/B$).
This term is the analog of (\ref{syserrW}) for the $\Gamma$-method.
We see that the integrated autocorrelation time is effectively constructed
by a double sum within one bin of size $B$ and the $\tau/B$ error
is a finite size effect which masks the actual $\exp(-B/\tau)$ expected
from truncation.
The many bins are only used to tame
statistical fluctuations of the estimate but do not improve the 
systematic error. The analogous finite size effect for the $\Gamma$-method
occurs only at order $\tau/N$.
A cancellation of the $\tau/B$ or $\exp(-W/\tau)$ 
effects dominant for the two methods was not found to be practical, as one
would have to use a noisy estimate for $\tau$ obtained from the data
and would not know the precise pre-factor.

In (\ref{sigmabin}) the function $f$ has to be evaluated on averages over only $B$ events, that is
only the $N_B$'th fraction of all measurements. This may be affected with stability
problems in the case of fits for example.
Compared to our earlier discussion, the gradient
of $f$ is here implicitly formed stochastically
with larger average spreads $\sim \sqrt{N_B} \, h_\alpha$. 
Both problems are cleverly overcome by going
to complementary or jackknife bins,
\be
c_\alpha^k = \frac1{N-B} \left( \sum_{i=1}^N a_\alpha^i -B\, b_\alpha^k \right).
\ee
An elementary calculation yields the relation
\be
\left\langle
(c_\alpha^k-\bar{a}_\alpha)(c_\beta^k-\bar{a}_\beta)
\right \rangle
=
\frac{1}{(N_B-1)^2}
\left\langle
(b_\alpha^k-\bar{a}_\alpha)(b_\beta^k-\bar{a}_\beta)
\right \rangle.
\ee
Hence
the resulting jackknife error estimator is
\be
\bar{\sigma}^2_{F\, {\rm jack}} 
= \frac{N_B-1}{N_B} \sum_{k=1}^{N_B} (f(c_\alpha^k) -\bar{F})^2 \, .
\label{sigmajack}
\ee
Here all evaluations of $f$ practically work on the full statistics,
and the spread of its arguments is of the same order $h_\alpha$ as
taken for the numerical derivative in the $\Gamma$-method.
The result of the above analysis of systematic errors is
however unchanged for the jackknife estimate.

For completeness we finally mention that also under binning the scatter between
$\bar{F}$, the function $f$ applied to the mean over all data, 
and the average of $f(c_\alpha^k)$ over all bins can be used to
cancel the leading $\rO(1/N)$ bias in the estimate of the mean of derived observables.
In analogy to (\ref{unbias}) we have to substitute
\be
\bar{F} \to N_B \, \bar{F} +(1-N_B)\,
\frac1{N_B} \sum_{k=1}^{N_B} f(c_\alpha^k) \, .
\ee

\section{An implementation of the $\Gamma$-method \label{appmatlab}}

In the ALPHA collaboration we have recently performed most data
analyses in Matlab, since we found that it combines comfortable programming,
a robust library, interactive graphics and acceptable speed suitable
for this purpose. 
An improved version of the well tested core-routine is described in the first subsection. 
While it has been applied to many realistic data sets in the ALPHA projects,
we here develop a simulator, that allows for clinical testing with
exactly known  errors in a situation that we consider representative for
simulations in lattice QCD.

\subsection{Description of the Matlab routine {\tt UWerr}}

The calling sequence of the function is
\begin{verbatim}
     [value,dvalue,ddvalue,tauint,dtauint,Q] = ...
     UWerr(Data,Stau,Nrep,Name,Quantity,P1,P2,...)
\end{verbatim}
with the meaning of the input and output arguments
described below. Similar information is obtained from {\tt help UWerr}
under Matlab.\\[2ex]
Input:\\[1ex]
\underline{{\tt Data}} 
is a matrix filled with the primary estimates $a_\alpha^{i,r}$
from $R$ replica with $N_1, N_2, \ldots, N_R$ measurements each,
\be
{\tt Data} = 
\left(\begin{array}{lllc}
a_1^{1,1} & a_2^{1,1} & a_3^{1,1} & \cdots \\[0.2ex]
a_1^{2,1} & a_2^{2,1} & a_3^{2,1} & \cdots \\
\vdots & \vdots & \vdots & \vdots \\
a_1^{N_1,1} & a_2^{N_1,1} & a_3^{N_1,1} & \cdots \\[0.2ex]
a_1^{1,2} & a_2^{1,2} & a_3^{1,2} & \cdots \\
\vdots & \vdots & \vdots & \vdots \\
a_1^{N_R,R} & a_2^{N_R,R} & a_3^{N_R,R} & \cdots \\
\end{array}\right)
\nonumber\ee
All input arguments that follow can be omitted or set to empty (=[]) and are
then assigned default values.

If data after reading them into Matlab occur in a different matrix format,
e.g. one long vector, the Matlab commands {\tt reshape} may be employed
to transform them very efficiently and safely.\\[0.5ex]
\underline{{\tt Stau}} is the estimate $S=\tau/\tau_{\rm int}$ explained in
section (\ref{window}). The default is $S=1.5$.
It is tried to determine an optimal window $W$ as described,
but it is never taken larger than $\min\{N_r\}/2$.
If the windowing condition cannot be met, a warning is printed.\\[0.5ex]
\underline{{\tt Nrep}} $=[N_1,N_2,\ldots,N_R]$ 
is a vector whose elements specify the lengths of the individual
replica.
As default all data are interpreted as one replicum, 
{\tt Nrep = [N]}. Even if only one run has been made, one may specify
{\tt Nrep} such as to analyze sections of the history independently
and investigate the scatter of sub-averages.\\[0.5ex]
\underline{{\tt Name}} can be a string (default: {\tt 'NoName'}) which is
used as a title in the generated plots (see below).
If the argument is given but is not a string (an integer for example)
no plots are generated.\\[0.5ex]
\underline{{\tt Quantity}} 
can either be an integer from the range of the index $\alpha$,
in which case the corresponding primary quantity is analyzed. Alternatively
it may be a function handle\footnote{
At present also a string with the name can be entered, which however is obsolete
and announced not to be supported in future versions of Matlab.
} 
{\tt @fun } if {\tt fun } is the name of
the function $f(A_1,A_2,\ldots)$ 
which has to be user-defined and has  as first argument 
a row-vector  which
has as many entries as {\tt Data} has columns. 
In this case the corresponding
derived quantity is analyzed. Optional further arguments $P1,P2,\ldots$
are passed on to the function as second, third, \ldots argument.
The default value is {\tt Quantity=1}.\\[2ex]
\noindent Output:\\[1ex]
\underline{{\tt value,dvalue,ddvalue,tauint,dtauint,Q}} have their obvious meaning
as best estimates of 
$F,\sigma_F$ as defined in (\ref{Fdef}),(\ref{sigmaF}), 
statistical error of the error from (\ref{Sokalformula}),
$\tauint{F}$ and $\delta(\tauint{F})$
given in (\ref{tauintF}) and extracted from (\ref{Sokalfortau}).
The $Q$-value (\ref{Qdef}) follows if $R \ge 2$.
Note that $C_F, v_F$ and the chosen 
$W$ can be reconstructed from the output if necessary.

The leading bias is corrected according to (\ref{unbias}) for $R \ge 2$.
If this correction amounts to more than $\sigma_F/4$ the user is warned.
The routine generates --- unless suppressed --- plots of 
our estimate for $\rho(t)=\Gamma_F(t)/\Gamma_F(0)$
and $\bbtauint{F}(W)$ with errorbars
in the relevant range to inspect 
the required plateau behaviour.
Errors of $\rho(t)$ have been added in version 6
of the {\tt UWerr}.
They are computed with formula (E.10)
in \cite{Luscher:2005rx} taking $\Lambda=t+W$.
If there are two or more replica their distribution (\ref{repdist}) 
is shown as a histogram with its $Q$-value (\ref{Qdef}) in the title.
In addition a histogram of all $a_\alpha^{i,r}$  
is displayed in case that a primary observable
is analyzed.

The code (m-file) of the Matlab routine {\tt UWerr.m} described here together with
a sample function {\tt effmass.m} used in the following test can be down-loaded
from {\tt www-com.physik.hu-berlin.de/ALPHAsoft}

\subsection{Test in a simulator}

\begin{figure}[tb]
\begin{center}
\epsfig{file=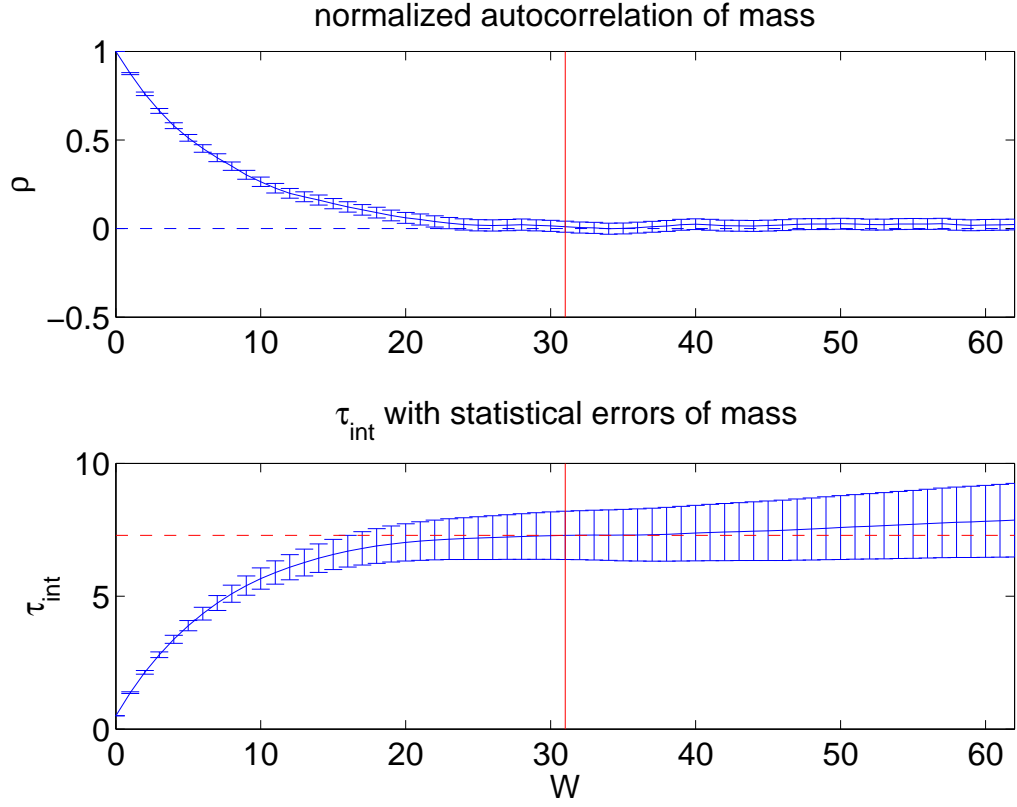,width=\textwidth}
\end{center}
\caption{Generated plots of $\bbar{\Gamma}_F(t)$ and $\bbtauint{F}(W)$
with a vertical line at the $W$-value picked automatically
and a horizontal one at the resulting
estimate of $\tauint{F}$.}
\label{auto}
\end{figure}
\begin{figure}[tb]
\begin{center}
\epsfig{file=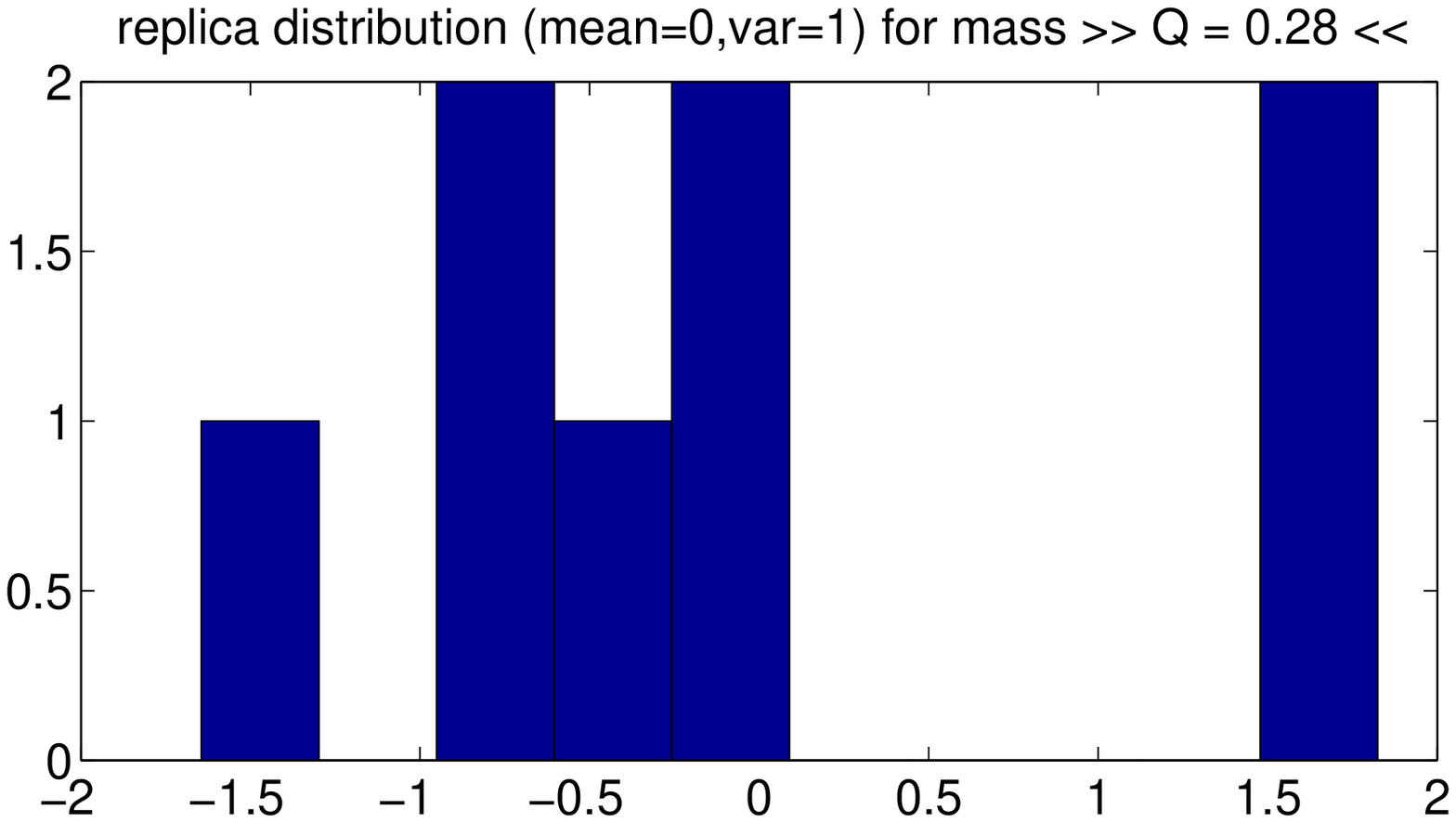,width=0.9\textwidth}
\end{center}
\caption{Replica distribution of $\{p_r\}$ with the $Q$-value in the title.}
\label{reps}
\end{figure}

As a test case
we think of determining the `effective' mass 
\be
m=\ln(G(z)/G(z+1)) = f(G(z),G(z+1)) \equiv F
\ee
from a correlation $G(z)=\exp(-mz)$ in euclidean time $z$,
which we take simply exponential since we are not
interested here in systematic errors of $m$.
To fake noisy (auto)correlated data for $G$ we employ
normally distributed random numbers $\eta_i, i=1,2,\ldots$ with
zero mean and unit variance to produce a
sequence $\{ \nu_i \}$ by recursively setting\footnote{
I thank Hubert Simma for pointing out
an improvement here.}
\be
\nu_1 = \eta_1, \quad \nu_{i+1} = \sqrt{1-a^2}\, \eta_{i+1} + a \nu_{i},
\ee
with
\be
\quad a = \frac{2\tau-1}{2\tau+1}, \quad \tau \ge \frac12.
\ee
A short calculation yields
\be
\langle\, \nu_{i} \,\rangle =0, \quad
\langle\, \nu_{i+t} \nu_{i} \,\rangle = a^t, \quad 
\sum_{t=-\infty}^{\infty} a^{|t|} = 2 \tau .
\ee
We use three independent such sequences $\nu^{(1)}_i, \nu^{(2)}_i,\nu^{(3)}_i$
and choose for them integrated autocorrelation times
$\tau^{(1)} < \tau^{(2)}=\tau^{(3)}$.
 
Synthetic data to measure $m = f(A_1,A_2)$ are formed,
\be
a^i_1 = G(0) + q (\nu^{(1)}_i + \nu^{(2)}_i), \quad 
a^i_2 = G(1) + q (\nu^{(1)}_i + \nu^{(3)}_i) \, .
\ee
The real parameter $q$ sets the strength of the noise, and the presence of $\nu^{(1)}_i$
in both observables creates a correlation between them, as
in real life.

All relevant quantities of the error analysis can be worked out exactly
for this case,
\be
v_{A_1} = v_{A_2} = 2 q^2, \quad
\tauint{A_1} = \tauint{A_2} = (\tau^{(1)}+\tau^{(2)})/2 ,
\ee
\be
v_F = 2 q^2 (1+\exp(2m)-\exp(m)) ,
\ee
\be
\tauint{F} = \frac{\hat{m}^2/2}{\hat{m}^2+1} \, \tau^{(1)}
+ \frac{\hat{m}^2/2+1}{\hat{m}^2+1} \, \tau^{(2)}
\ee
with
\be
\hat{m} = 2 \sinh(m/2).
\ee
Note that in the weighed sum giving $\tauint{F}$ the time scale
$\tau^{(1)}$ contributes with a small weight, due to correlations
between $a_i^1, a_i^2$  and the cancellation of the corresponding fluctuations
in the effective mass.

For our test we now set $m=0.2, \tau^{(1)}=4, \tau^{(2)}=8, q=0.2$ 
which leads to
$v_F =0.1016, \tauint{F}=7.92$. We generate $8\times 1000$ estimates
corresponding to an error in the mass of 0.0142  and get
\begin{verbatim}
>> Nr=ones(8,1)*(N/8);
>> [value,dvalue,ddvalue,tauint,dtauint,Q]= ...
   UWerr(Data,1,Nr,'mass',@effmass,1,2);
>> [value,dvalue,ddvalue,tauint,dtauint,Q]
ans =

    0.2128    0.0134    0.0008    7.2909    0.7396   0.2827
\end{verbatim}
The generated plots are shown in Fig.\ref{auto} and in Fig.\ref{reps}.
By repeating such an analysis 20000 times we knock down the statistical
error of the error far enough to exhibit a systematic error of
$\sim  -0.5 \%$ which is of the expected order $\frac12 \exp(-W/\tau)$
with $W=37$ and $\tau = 8$, while, of course, the actual automatic $W$
and estimated $\tau$-values fluctuate.

\begin{figure}[tb]
\begin{center}
\epsfig{file=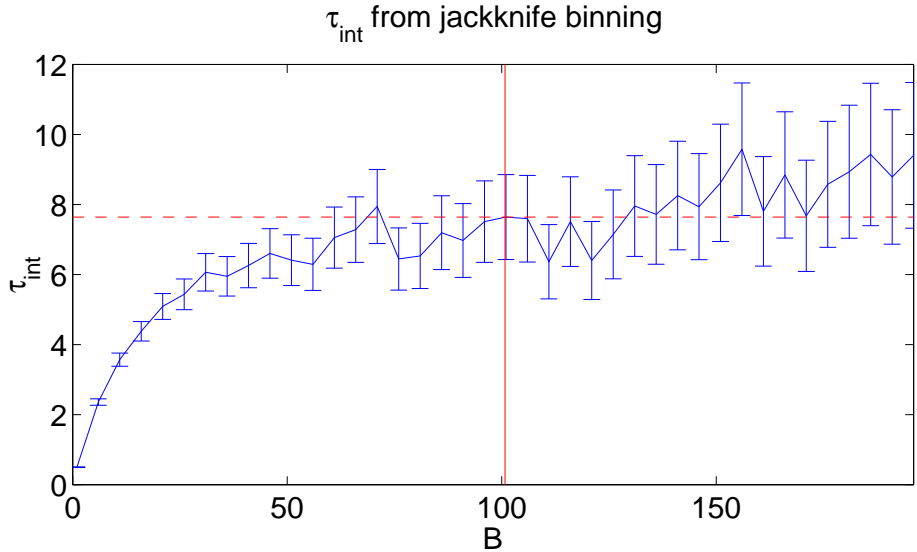,width=0.9\textwidth}
\end{center}
\caption{Estimation of $\tauint{F}$ by jackknife binning.
A vertical line shows the optimal bin size (\ref{Bopt}).}
\label{jack}
\end{figure}
The jackknife analog of the lower part of Fig.\ref{auto} can be seen
in Fig.\ref{jack} for the same data. 
It looks consistent with a less pronounced plateau,
as expected.

More relevant are tests on real data. The routine passed a number of such tests,
but, as usual with software, with probability one some future application
will force further evolution.

\end{appendix}

\bibliography{refs}           
\bibliographystyle{h-elsevier}   

\end{document}